\documentclass[aps,prd,twocolumn,showpacs,amsmath,amssymb]{revtex4-1}
\usepackage{amsmath} \usepackage{graphicx} \usepackage{subfigure}
\usepackage{epstopdf} \usepackage{color} \usepackage{multirow}
\usepackage{setspace} \usepackage{overpic} \usepackage{amssymb}
\usepackage[bookmarksnumbered, pdfstartview=FitH,colorlinks,urlcolor=blue, citecolor=blue,linkcolor=blue] {hyperref}
\usepackage{lineno}
\usepackage{bm}
\usepackage{rotating}
\usepackage[utf8]{inputenc}

\let\oldequation\equation
\let\oldendequation\endequation
\renewenvironment{equation}
 {\linenomathNonumbers\oldequation}
 {\oldendequation\endlinenomath}

\begin{document}


\title{\boldmath Observation of $\chi_{cJ}\to \Lambda\bar \Lambda \eta$}

\author{
M.~Ablikim$^{1}$, M.~N.~Achasov$^{11,b}$, P.~Adlarson$^{70}$, M.~Albrecht$^{4}$, R.~Aliberti$^{31}$, A.~Amoroso$^{69A,69C}$, M.~R.~An$^{35}$, Q.~An$^{66,53}$, X.~H.~Bai$^{61}$, Y.~Bai$^{52}$
% ...
\\
\vspace{0.2cm}
(BESIII Collaboration)\\
\vspace{0.2cm} {\it
$^{1}$ Institute of High Energy Physics, Beijing 100049, People's Republic of China\\
% ...
$^{4}$ Bochum Ruhr-University, D-44780 Bochum, Germany\\
% ...
$^{11}$ G.I. Budker Institute of Nuclear Physics SB RAS (BINP), Novosibirsk 630090, Russia\\
% ...
$^{31}$ Johannes Gutenberg University of Mainz, Johann-Joachim-Becher-Weg 45, D-55099 Mainz, Germany\\
% ...
$^{35}$ Liaoning Normal University, Dalian 116029, People's Republic of China\\
% ...
$^{52}$ Southeast University, Nanjing 211100, People's Republic of China\\
$^{53}$ State Key Laboratory of Particle Detection and Electronics, Beijing 100049, Hefei 230026, People's Republic of China\\
% ...
$^{61}$ University of Jinan, Jinan 250022, People's Republic of China\\
% ...
$^{66}$ University of Science and Technology of China, Hefei 230026, People's Republic of China\\
% ...
$^{69}$ University of Turin and INFN, (A)University of Turin, I-10125, Turin, Italy; (B)University of Eastern Piedmont, I-15121, Alessandria, Italy; (C)INFN, I-10125, Turin, Italy\\
$^{70}$ Uppsala University, Box 516, SE-75120 Uppsala, Sweden\\
% ...
$^{b}$ Also at the Novosibirsk State University, Novosibirsk, 630090, Russia\\
% ...
}
\vspace{0.4cm}
}

\begin{abstract}
  By analyzing $(448.1\pm2.9)\times10^6$ $\psi(3686)$ events collected
  with the BESIII detector operating at the BEPCII collider, the
  decays of $\chi_{cJ} \to \Lambda\bar \Lambda \eta$ ($J=0$, 1 and 2)
  are observed for the first time with statistical significances of
  $13.9\sigma$, $6.7\sigma$, and $8.2\sigma$, respectively.  The
  product branching fractions of $\psi(3686)\to\gamma\chi_{cJ}$ and
  $\chi_{cJ}\to \Lambda\bar \Lambda \eta$ are measured.  Dividing by
  the world averages of the branching fractions of
  $\psi(3686)\to\gamma\chi_{cJ}$, the branching fractions of
  $\chi_{cJ}\to \Lambda\bar \Lambda \eta$ decays are determined to be
  $(2.31\pm0.30\pm0.21)\times10^{-4}$,
  $(5.86\pm1.38\pm0.68)\times10^{-5}$, and
  $(1.05\pm0.21\pm0.15)\times10^{-4}$ for $J=0$, 1 and 2,
  respectively, where the first uncertainties are statistical and the
  second systematic.
\end{abstract}


\maketitle

\section{Introduction}


Studies of the processes involving $B\bar{B}P$, where $B$ and $P$
denote baryons and pseudoscalar mesons, respectively, are important to
search for possible $B\bar B$ threshold enhancements and excited
baryon states decaying into $BP$.
An enhancement around the $\Lambda\bar{\Lambda}$ production threshold was
observed in the $e^+e^-\to \phi\Lambda\bar{\Lambda}$
process~\cite{LLbarphi}, and the interpretation of
$\Lambda\bar{\Lambda}$ enhancement originating from decay of the
$\eta(2225)\to\Lambda\bar{\Lambda}$~\cite{LLbar_theory1} was rejected
with a significance of 7$\sigma$. Similar structure was also reported in the $B$ meson decays $B^0\to\Lambda\bar{\Lambda}K^0$ and $B^+\to\Lambda\bar{\Lambda}K^+$~\cite{B2LLK}. On the other hand, an excited
$\Lambda$ state, $\Lambda(1670)$, was observed in the $\Lambda \eta$
mass spectra in the near-threshold reaction $K^-p\to
\eta\Lambda$~\cite{Leta} and the charmonium decay $\psi(3686)\to
\Lambda\bar \Lambda \eta$~\cite{abc}.  However, experimental results
on the $\Lambda\bar{\Lambda}$ production threshold enhancement and on
excited $\Lambda$ states decaying into $\Lambda \eta$ are still
limited. Comprehensive investigations of the $B\bar{B}P$ system in the
various charmonium state decays are desirable.  To date, only a few
studies of $\chi_{cJ}\to B\bar{B}P$ ($J=0,1,2$) have been
performed~\cite{pdg2020}, and no investigation of
$\chi_{cJ}\to\Lambda\bar{\Lambda}\eta$ has been reported.  Observation
of $\chi_{cJ}\to\Lambda\bar{\Lambda}\eta$ would provide an opportunity
to better understand the enhancement around the $\Lambda\bar \Lambda$
production threshold and a possible excited $\Lambda$ state, e.g.,
$\Lambda(1670)$.

In the quark model, the $\chi_{cJ}$ mesons are identified
as $^3P_J$ charmonium states.  Because of parity conservation, they
can not be produced by $e^+e^-$ annihilation directly.  As a result,
the decays of $\chi_{cJ}$ have not been studied as extensively as the
vector charmonium states $J/\psi$ and $\psi(3686)$ in both experiment
and theory.  However, the radiative decays of $\psi(3686)$ into
$\chi_{cJ}$ mesons have branching fractions of about
9\%~\cite{pdg2020} for each $\chi_{cJ}$ state, thereby offering an
ideal testbed to investigate $\chi_{cJ}$ meson decays.

In this paper, by analyzing $(448.1\pm2.9)\times 10^6$ $\psi(3686)$
events~\cite{ref::psip-num-inc} collected with the BESIII
detector~\cite{Ablikim:2009aa}, we present the first measurements of
the branching fractions of $\chi_{cJ}$ decays to $\Lambda\bar \Lambda
\eta$.

\section{BESIII DETECTOR AND MONTE CARLO SIMULATION}
\label{sec:BES}

The BESIII detector is a magnetic spectrometer~\cite{Ablikim:2009aa}
located at the Beijing Electron Positron
Collider~(BEPCII)~\cite{Yu:IPAC2016-TUYA01}.  The cylindrical core of
the BESIII detector consists of a main drift chamber filled with
helium-based gas~(MDC), a plastic scintillator time-of-flight
system~(TOF), and a CsI(Tl) electromagnetic calorimeter~(EMC), which
are all enclosed in a superconducting solenoidal magnet providing a
1.0~T magnetic field. The flux-return yoke is instrumented with
resistive plate chambers arranged in 9 layers in the barrel and 8
layers in the endcaps for muon identification. The acceptance of
charged particles and photons is 93\% of $4\pi$ solid angle.  The
charged-particle momentum resolution at 1.0~GeV/$c$ is $0.5\%$, and
the specific energy loss resolution is $6\%$ for the electrons from
Bhabha scattering. The EMC measures photon energies with a resolution
of $2.5\%$~($5\%$) at $1$~GeV in the barrel (end cap) region. The time
resolution of the TOF barrel part is 68~ps, while that of the end cap
part is 110~ps.

Simulated samples produced with the {\sc geant4}-based~\cite{GEANT4}
Monte-Carlo~(MC) software, which includes the geometric description of
the BESIII detector and the detector response, are used to determine
the detection efficiencies and to estimate the background levels. The
simulation takes into account the beam energy spread and initial state
radiation~(ISR) in the $e^+e^-$ annihilation modeled with the
generator {\sc kkmc}~\cite{KKMC}. The inclusive MC samples consist of
$5.06\times10^{8}$ $\psi(3686)$ events, the ISR production of the
$J/\psi$ state, and the continuum processes incorporated in {\sc
  kkmc}. The known decay modes are modeled with {\sc
  evtgen}~\cite{evtgen} using the branching fractions taken from the
Particle Data Group~\cite{pdg2020}, and the remaining unknown decays
from the charmonium states with {\sc lundcharm}~\cite{LUNDCHARM1}.
Final state radiation is incorporated with {\sc photos}~\cite{PHOTOS}.

For each signal process of $\psi(3686)\to\gamma\chi_{cJ}, \chi_{cJ}\to \Lambda\bar \Lambda \eta, \Lambda(\bar{\Lambda})\to p\pi^-(\bar{p}\pi^+)$, $5\times10^5$ signal MC
events are generated. Radiative decay $\psi(3686)\to\gamma\chi_{cJ}$ is generated with a $1+\lambda\cos^2\theta$ distribution,
where $\theta$ is the angle between the direction of the radiative
photon and the beam, and $\lambda=1,-1/3,1/13$ for $J=0,1,2$~\cite{evtgen} in
accordance with expectations for electric dipole transitions.
Intrinsic width and mass values as given in Ref.~\cite{pdg2020} are
used for the $\chi_{cJ}$ states in the simulation. The process of $\chi_{cJ}\to \Lambda\bar \Lambda \eta$ is simulated with the BODY3 generator based on {\sc evtgen}~\cite{evtgen}, as discussed in section \ref{sec:mc}. The decay of $\Lambda(\bar{\Lambda})\to p\pi^-(\bar{p}\pi^+)$ is simulated in phase space.

\section{EVENT SELECTION}
\label{sec:selection}

In this analysis, the $\Lambda$, $\bar \Lambda$, and $\eta$ particles
are reconstructed via the $\Lambda \to p\pi^-$, $\bar \Lambda \to \bar
p\pi^+$, and $\eta\to \gamma\gamma$ decays, respectively.

\begin{figure}[htbp]
  \centering
  \includegraphics[width=\columnwidth]{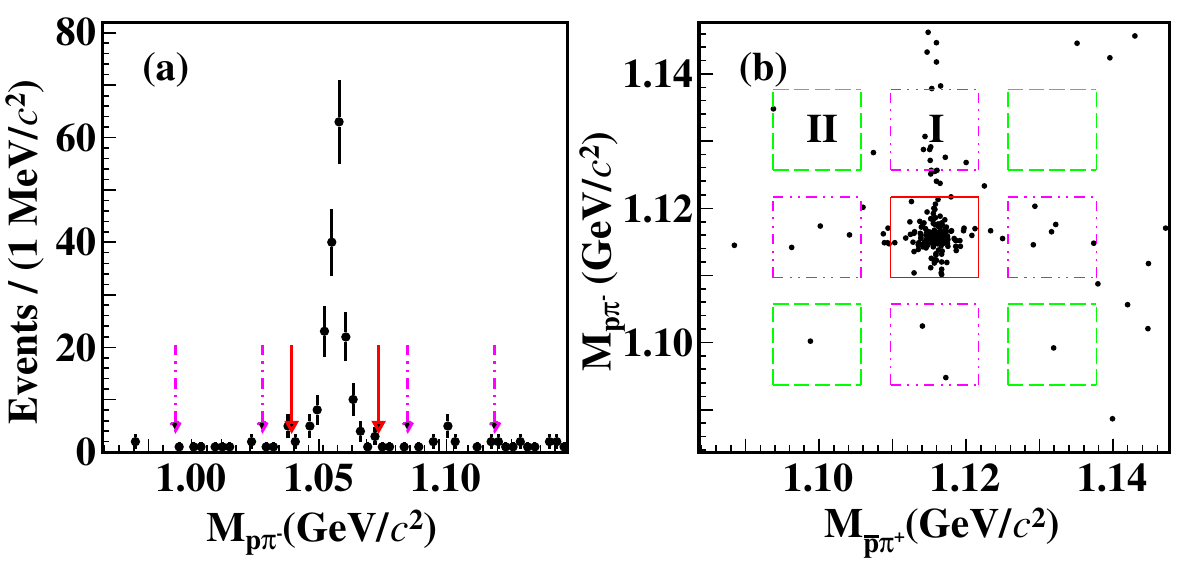}
  \caption{The distributions of (a) $M_{p\pi^-}$ and (b) $M_{p\pi^-}$
    versus $M_{\bar p\pi^+}$ of the candidates for $\psi(3686)\to
    \gamma\chi_{cJ}$ with $\chi_{cJ}\to \Lambda\bar \Lambda\eta$ in
    data, where all requirements except for the
    $\Lambda(\bar{\Lambda})$ signal region have been imposed.  In (a),
    the pair of red solid arrows denote the $\Lambda$ signal region,
    and the pairs of pink dot-dashed arrows denote sideband regions of
    the accepted candidates.  In (b), the red solid rectangle denotes
    the $\Lambda\bar \Lambda$ signal region, the pink dot-dashed
    rectangles denote the $\Lambda\bar \Lambda$ sideband
    \uppercase\expandafter{\romannumeral 1} region, and the green
    dashed rectangles denote the $\Lambda\bar \Lambda$ sideband
    \uppercase\expandafter{\romannumeral 2} region.  }
  \label{fig:lmd}
\end{figure}

All charged tracks are required to satisfy $|Z_v|<20$ cm and $|\rm
\cos\theta|<0.93$, where $Z_v$ denotes the distance from the
interaction point to the point of closest approach of the
reconstructed track to the $z$ axis, which is the symmetry axis of the
MDC, and $\theta$ is the polar angle relative to the $z$ axis.
Candidate events must have four charged tracks with zero net charge
and at least three good photons. The $\Lambda(\bar \Lambda)$
candidates are reconstructed using vertex fits of all oppositely
charged track pairs, which are assumed to be $p\pi^-(\bar p\pi^+)$
without particle identification. To suppress the $p\pi^-(\bar p\pi^+)$
combinatorial background, the reconstructed decay lengths of the
$\Lambda(\bar \Lambda)$ candidates are required to be more than twice
their standard deviations. Figure~\ref{fig:lmd} shows the distributions of $M_{p\pi^-}$ and $M_{p\pi^-}$ versus $M_{\bar{p}\pi^+}$ of survived candidates in data. The invariant mass of $p\pi^-(\bar
p\pi^+)$ is required to be within the $\Lambda(\bar \Lambda)$ signal
region, defined as
$|M_{p\pi^-(\bar{p}\pi^+)}-m_{\Lambda(\bar{\Lambda})}|< 6$\,MeV/$c^2$,
where $m_{\Lambda(\bar{\Lambda})}$ is the $\Lambda(\bar \Lambda)$
nominal mass~\cite{pdg2020}, while the one-dimensional (1D)
$\Lambda(\bar \Lambda)$ sideband region is defined as
$10<|M_{p\pi^-(\bar{p}\pi^+)}-m_{\Lambda(\bar{\Lambda})}|<
22$\,MeV/$c^2$. The two-dimensional (2D) $\Lambda\bar \Lambda$ signal
region is defined as the square region with both $p\pi^-$ and $\bar
p\pi^+$ combinations lying in the $\Lambda(\bar \Lambda)$ signal
regions. The $\Lambda\bar \Lambda$ sideband
\uppercase\expandafter{\romannumeral 1} regions are defined as the
square regions with either one of the $p\pi^-$ or $\bar p\pi^+$
combinations locating in the 1D $\Lambda(\bar \Lambda)$ sideband
regions and the other in the 1D signal region. The sideband
\uppercase\expandafter{\romannumeral 2} regions are defined as the
square regions with both $p\pi^-$ and $\bar p\pi^+$ combinations
locating in the 1D $\Lambda(\bar \Lambda)$ sideband regions.

Good photon candidates are chosen from isolated clusters in the
EMC. Their energies are required to be greater than 25 MeV in the
barrel ($|\cos\theta|<0.8$) and 50 MeV in the end cap
($0.86<|\cos\theta|<0.92$) regions. Reconstructed clusters due to
electronic noise or beam backgrounds are suppressed by requiring the
timing information to be within [0, 700] ns after the event start
time. To suppress fake photons produced by hadronic interactions in
the EMC and secondary photons from bremsstrahlung, clusters within a
cone angle of $20^\circ$ around the extrapolated position in the EMC
of any charged track are rejected. The energy deposited in the
neighbor TOF counters is taken into account to improve the
reconstruction efficiency and energy resolution.

To further suppress the combinatorial background, a four-momentum
conservation constraint (4C) kinematic fit under the hypothesis of
$e^+e^-\to \Lambda\bar \Lambda\gamma\gamma\gamma$ is applied to the
events. The combination with the minimum $\chi^2_{\rm 4C}$ is kept for
further analysis. The $\chi_{\rm 4C}^{2}$ of the kinematic fit is
required to be less than 50 based on optimization with the Punzi significance method~\cite{punzi_opt} with the formula $\frac{\epsilon}{1.5+\sqrt{B}}$, where $\epsilon$ is the detection efficiency and $B$ is the number of background events from the inclusive $\psi(3686)$ MC sample. This requirement will reject 84\% of background and lose 14\% of signal efficiency.

\begin{figure}[htbp]
	\centering
	\includegraphics[width=\columnwidth]{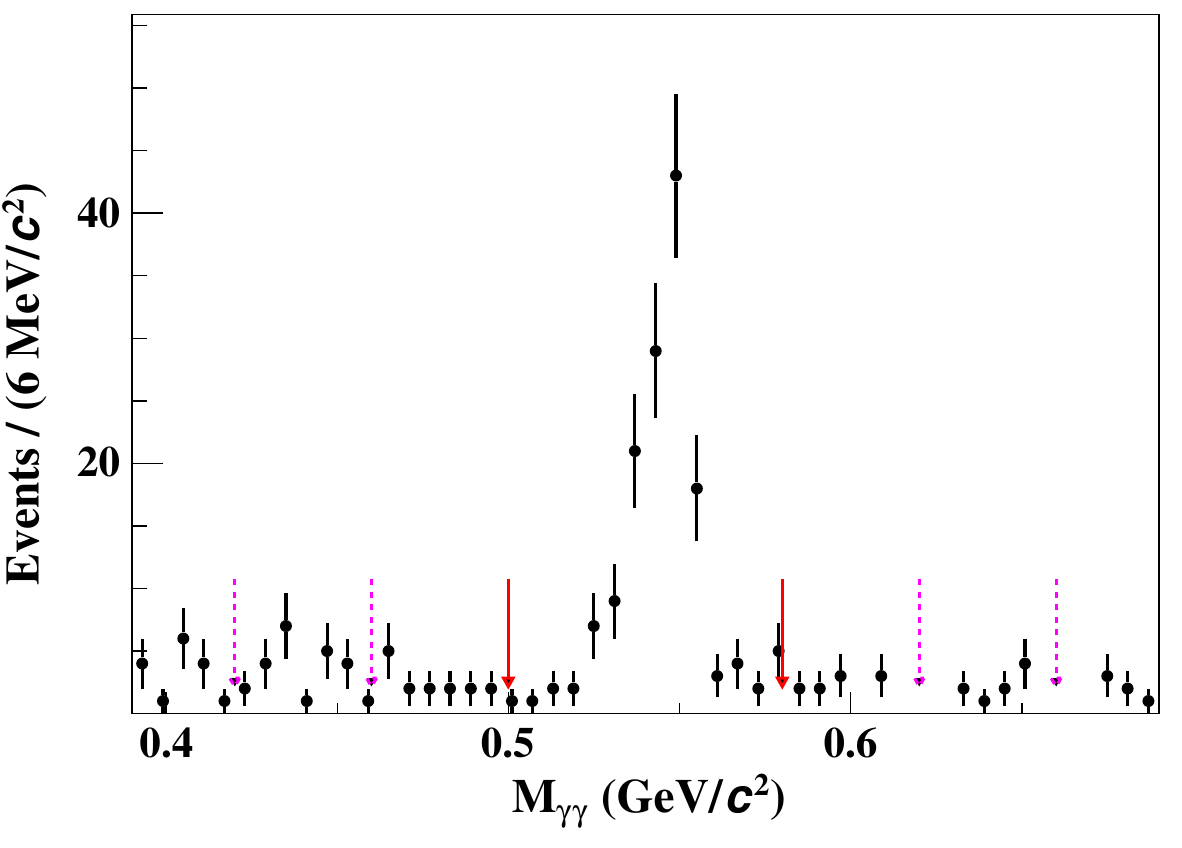}
	\caption{The $M_{\gamma\gamma}$ distributions of $\eta$ candidates of the accepted events in data. The pair of red solid arrows denote the $\eta$ signal region and the pairs of pink dot-dashed arrows denote the $\eta$ sideband regions.}
	\label{fig:eta}
\end{figure}

The three selected photons are ordered according to their energies with
$E_{\gamma_{1}}>E_{\gamma_{2}}>E_{\gamma_{3}}$, defined as $\gamma_{1}$, $\gamma_{2}$, and $\gamma_{3}$. The $\eta$ candidates are reconstructed from
either $\gamma_1\gamma_2$ or $\gamma_1\gamma_3$ pair.
Based on MC study, excluding $\eta$ candidates from $\gamma_{2}\gamma_{3}$ could suppress the background by 6\% with the signal efficiency loss less than 0.1\%, and improves the Punzi significance by 3\%.  Figure~\ref{fig:eta} shows the $M_{\gamma\gamma}$ distribution of $\eta$ candidates of the accepted events in data. The $\eta$ signal and
sideband regions are defined as
$|M_{\gamma\gamma}-0.54|<0.04$\,GeV/$c^2$ and
$0.08<|M_{\gamma\gamma}-0.54|<0.12$\,GeV/$c^2$, respectively.
Events with both $M_{\gamma_{1}\gamma_{2}}$ and $M_{\gamma_{1}\gamma_{3}}$ being in the
$\eta$ signal region are excluded from analysis.
Events with neither $M_{\gamma_1\gamma_2}$ or $M_{\gamma_1\gamma_3}$ within the signal region,
but either $M_{\gamma_1\gamma_2}$ or $M_{\gamma_1\gamma_3}$ in the
sideband region, are taken as sideband events.

To suppress background events associated with $J/\psi\to\Lambda\bar{\Lambda}\gamma$, events with
$|M_{\Lambda\bar{\Lambda}\gamma_{1,2,3}}-3.101|<0.044$\,GeV/$c^2$ are
vetoed, where the veto region is five standard deviations of the
resolution around the mean value. 
To suppress background events related to $\chi_{cJ}\to\Lambda\bar{\Lambda}\pi^0$, events with
$|M_{\gamma\gamma}-0.137|<0.015$\,GeV/$c^2$ are excluded for all three
possible photon pairs. To suppress background events associated with
$\Sigma(\bar{\Sigma})\to\Lambda(\bar{\Lambda})\gamma$, events with
$|M_{\Lambda(\bar{\Lambda})\gamma_{2,3}}-1.192|<0.015$\,GeV/$c^2$ are
rejected.

\begin{figure*}[htbp]
	\centering
	\includegraphics[width= 17cm]{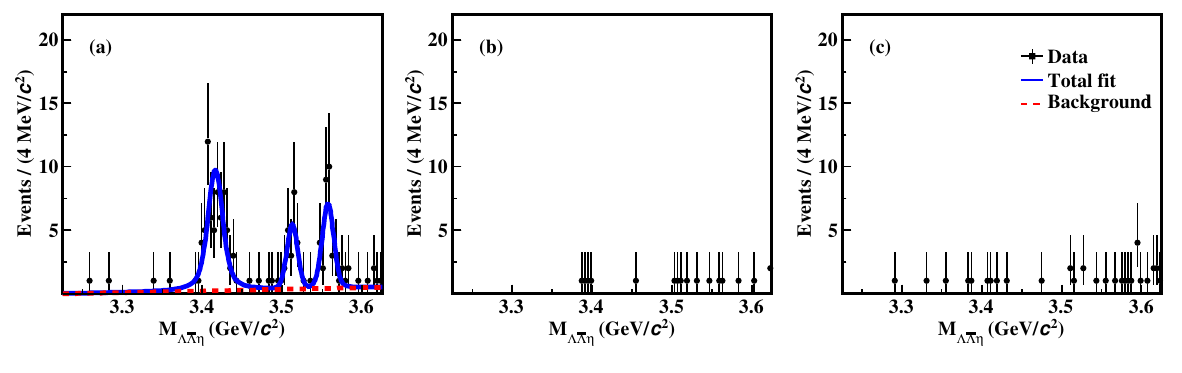}
	\caption{The $M_{\Lambda\bar{\Lambda}\eta}$ distributions of the accepted events in (a) the combined $\Lambda\bar{\Lambda}$ and $\eta$ signal region, (b) the $\Lambda\bar{\Lambda}$ sideband \uppercase\expandafter{\romannumeral 1} region, and (c) the $\eta$ sideband region. The points with error bars are data, the red dashed lines are the fitted background shapes, and the blue solid lines are the overall fits.}
	\label{fig:fit}
\end{figure*}

\begin{figure*}[htbp]
	\centering
	\includegraphics[width=17cm]{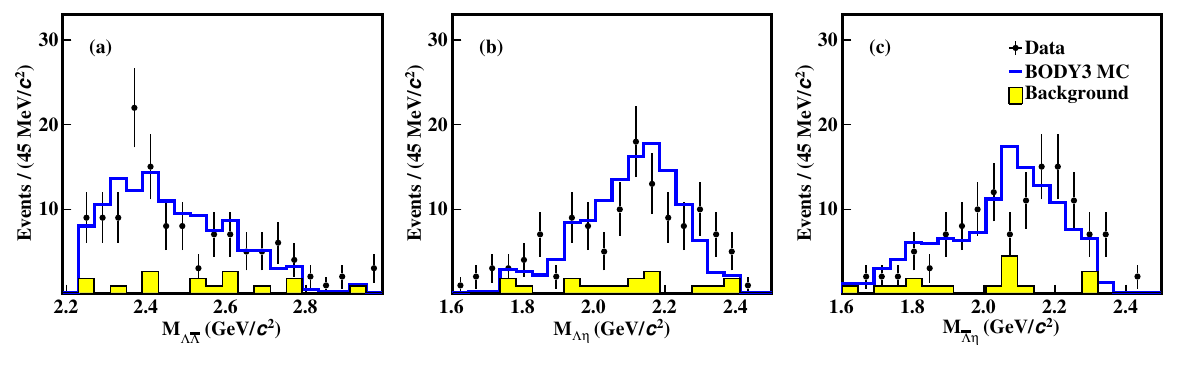}
	\caption{The distributions of (a) $M_{\Lambda\bar{\Lambda}}$,
		(b) $M_{\Lambda\eta}$, and (c) $M_{\bar{\Lambda}\eta}$ from
		the three $\chi_{cJ}$ signal regions of the data sample. The
		dots with error bars are data. The blue solid curves are
		BODY3 MC events. The yellow hatched histograms are the
		simulated backgrounds. The signal and background yields have
		been normalized to the statistics of data.}
	\label{fig:resonance}
\end{figure*}

A total of 144 candidate events survive from all requirements.
Figure~\ref{fig:fit}(a) shows the distribution of $M_{\Lambda\bar \Lambda\eta}$ of the accepted candidate events in data. Clear signals of
$\chi_{c0}$, $\chi_{c1}$, and $\chi_{c2}$ are observed.  The
distributions of $M_{\Lambda\bar{\Lambda}}$, $M_{\Lambda\eta}$, and
$M_{\bar{\Lambda}\eta}$ from all $\chi_{cJ}$ signal regions of the
data sample are shown in Fig.~\ref{fig:resonance}. Here, the signal
regions of $\chi_{c0}$, $\chi_{c1}$, and $\chi_{c2}$ are defined as
[3.385, 3.445]\,GeV/$c^2$, [3.490, 3.530]\,GeV/$c^2$, and [3.536,
  3.576]\,GeV/$c^2$, respectively. With present statistics, it is impossible to conclude that there is an enhancement near the
$\Lambda\bar{\Lambda}$ production threshold in
Fig.~\ref{fig:resonance}(a). In addition, no obvious excited $\Lambda$ state is
found in Fig.~\ref{fig:resonance}(b) or Fig.~\ref{fig:resonance}(c).
Meanwhile, we can not conclude whether there is any structure difference between the $M_{\Lambda\bar{\Lambda}}$, $M_{\Lambda\eta}$, and
$M_{\bar{\Lambda}\eta}$ spectra from different $\chi_{cJ}$ signal regions.

\section{BACKGROUND STUDIES}
\label{sec:background}

Possible non-$\Lambda(\bar{\Lambda})$ background and non-$\eta$ peaking background from
$\psi(3686)$ decays are studied with sideband
events. Figures~\ref{fig:fit}(b) and \ref{fig:fit}(c) show the $M_{\Lambda \bar{\Lambda} \eta}$ distributions of candidate events in the $\Lambda\bar{\Lambda}$ sideband \uppercase\expandafter{\romannumeral 1} region and $\eta$ sideband region, respectively. No significant non-$\Lambda(\bar{\Lambda})$ peaking background and non-$\eta$ peaking background is observed. For the $\Lambda\bar{\Lambda}$ sideband \uppercase\expandafter{\romannumeral 2} region, only three events are remained, which are negligible.

Potential backgrounds with $\Lambda\bar{\Lambda}\eta+X$ final states are estimated by analyzing the inclusive $\psi(3686)$ MC sample with
TopoAna~\cite{TopoAna}. No peaking background is found except for
$\chi_{c2}\to\Sigma^0\bar{\Lambda}\eta$. However, the
$\chi_{c2}\to\Sigma^0\bar{\Lambda}\eta$ decay is an isospin-violating process and no branching fraction is available.
The yield of this background is estimated by assuming that the ratio $\frac{\mathcal{B}(\chi_{c2}\to\Sigma^0\bar{\Lambda}\eta)}{\mathcal{B}(\chi_{c2}\to\Lambda\bar{\Lambda}\eta)}$
is comparable with
$\frac{\mathcal{B}(J/\psi\to\Sigma^0\bar{\Lambda})}{\mathcal{B}(J/\psi\to\Lambda\bar{\Lambda})}=1.5\%$
or
$\frac{\mathcal{B}(\psi(3686)\to\Sigma^0\bar{\Lambda})}{\mathcal{B}(\psi(3686)\to\Lambda\bar{\Lambda})}=3.2\%$~\cite{pdg2020} based on isospin symmetry.
MC studies show that the ratio of the yield of this background relative to our signal is less than $0.1\%$. Therefore this background is also negligible in this analysis.

Finally, the possible quantum electrodynamics (QED) contribution is
examined by using the continuum data corresponding to an integrated
luminosity of 44.45~pb$^{-1}$ taken at the center-of-mass energy of
3.65~GeV~\cite{lum}.  No event survives the selection criteria.
Therefore, the QED contribution is also neglected in this analysis.

\section{BRANCHING FRACTIONS}
\label{sec:mc}

To determine signal yields, an unbinned maximum likelihood fit is
performed on the $M_{\Lambda\bar \Lambda \eta}$ distribution of the accepted candidates in data.  In the
fit, the $\chi_{cJ}$ signals are described with individual
Breit-Wigner functions $\frac{1}{(M_{\Lambda\bar \Lambda \eta}-m_{\chi_{cJ}})^2+\Gamma^2_{\chi_{cJ}}/4}$ convolved with a Gaussian function. The masses
and widths of Breit-Wigner functions are fixed to their world average
values~\cite{pdg2020}. The mean and width of the Gaussian function are free
parameters.  Because the potential peaking background is negligible, a
linear function is chosen to describe the combinatorial background
shape.
For each signal decay mode, the statistical significance is calculated with $\Delta (\ln {\mathcal L})=\ln{\mathcal L}_{\rm max}-\ln{\mathcal L}_{\rm 0}$ and $\Delta ndf=3$. Here, the ${\mathcal L}_{\rm max}$ and ${\mathcal L}_0$ are the maximum likelihoods with and without the signal component in the fits; the $\Delta ndf$ is the variation of number of degrees of freedom.
The statistical signficances are $13.7\sigma$, $6.2\sigma$, and
$7.7\sigma$ for $\chi_{c0}\to \Lambda\bar{\Lambda}\eta$, $\chi_{c1}\to
\Lambda\bar{\Lambda}\eta$, and $\chi_{c2}\to
\Lambda\bar{\Lambda}\eta$, respectively. As shown in Fig.~\ref{fig:fit}(a), we obtain the
signal yields of $\chi_{c0}$, $\chi_{c1}$ and $\chi_{c2}$ to be
$66.9\pm8.8$, $21.3\pm5.0$, and $31.6\pm6.2$, respectively, where the
uncertainties are statistical only.

The $\chi_{cJ}\to \Lambda\bar \Lambda \eta$ decays are simulated with
a modified data-driven generator BODY3, which was developed to simulate different
intermediate states in data for a given three-body final
state. Initially, a phase space MC sample is used to determine
efficiency over the whole allowed kinematic region. Then, the Dalitz
plot of $M^2_{\Lambda \eta}$ versus $M^2_{\bar \Lambda \eta}$,
corrected for backgrounds and efficiencies, is used to determine the
probability of an event configuration generated randomly.
The detection efficiencies for
$\psi(3686)\to\gamma\chi_{c0,1,2}$ with $\chi_{c0,1,2}\to \Lambda\bar \Lambda \eta$ are $(4.11 \pm 0.03)\%$, $(5.17 \pm 0.03)\%$, and $(4.37 \pm 0.03)\%$, respectively, where the errors are statistical only.

The product branching fractions of $\psi(3686)\to\gamma\chi_{cJ}$ and $\chi_{cJ}\to \Lambda\bar \Lambda \eta$ are calculated with
\begin{equation}
	\begin{split}
	&\mathcal{B}({\psi(3686)\to\gamma\chi_{cJ}})\cdot\mathcal{B}({\chi_{cJ}\to \Lambda\bar \Lambda \eta})\\
	&=\frac{N_{\rm obs}^{ J}}{N_{\psi(3686)}\cdot\mathcal{B}^{2}({\Lambda \to p\pi^{-}})\cdot\mathcal{B}({\eta\to\gamma\gamma})\cdot\epsilon({\chi_{cJ}\to \Lambda\bar \Lambda \eta})},
	\end{split}
\end{equation}
where $N_{\rm obs}^{J}$ is the signal yield obtained from the fit to
the $M_{\Lambda\bar \Lambda \eta}$ distribution for $\chi_{cJ}$,
$N_{\psi(3686)}=(448.1\pm2.9)\times10^6$ is the number of
$\psi(3686)$ events~\cite{ref::psip-num-inc}, $\epsilon({\chi_{cJ}\to \Lambda\bar \Lambda \eta})$ is the detection efficiency for
$\chi_{cJ}$, and $\mathcal{B}({\Lambda \to p\pi^-})$ and
$\mathcal{B}({\eta\to\gamma\gamma})$ are the branching fractions of
$\Lambda \to p\pi^-$ and $\eta\to\gamma\gamma$ from
Ref.~\cite{pdg2020}.  Dividing by the world averages~\cite{pdg2020} of $\mathcal{B}({\psi(3686)\to\gamma\chi_{c0}})=(9.79\pm0.20)\%$, $\mathcal{B}({\psi(3686)\to\gamma\chi_{c1}})=(9.75\pm0.24)\%$, and $\mathcal{B}({\psi(3686)\to\gamma\chi_{c2}})=(9.52\pm0.20)\%$, we obtain the
branching fractions of $\chi_{cJ}\to \Lambda\bar \Lambda \eta$ decays. The obtained results are summarized in Table \ref{tab:Branching}.

\begin{table*}[htbp]
\centering\small
\caption{
Signal yields in data, detection efficiencies, and the branching fractions $\mathcal{B}({\psi(3686)\to\gamma\chi_{cJ}})$, $\mathcal{B}({\psi(3686)\to\gamma\chi_{cJ}})\cdot \mathcal{B}({\chi_{cJ}\to \Lambda \bar{\Lambda} \eta})$, and $\mathcal{B}({\chi_{cJ}\to \Lambda \bar{\Lambda} \eta})$. The first errors are statistical and the second systematic.}\label{tab:Branching}
\begin{tabular}{l ccc}
\hline
 \hline
                   & $\chi_{c0}$& $\chi_{c1}$ & $\chi_{c2}$                 \\  \hline
$N_{\rm obs}^{J}$  & $66.9\pm8.8$                 & $21.3\pm5.0$                 & $31.6\pm6.2$       \\
  $\epsilon(\chi_{cJ}\to \Lambda\bar \Lambda \eta)$     & $(4.11\pm0.03)\%$          & $(5.17\pm0.03)\%$          & $(4.37\pm0.03)\%$      \\
  $\mathcal{B}({\psi(3686)\to\gamma\chi_{cJ}})\cdot \mathcal{B}({\chi_{cJ}\to \Lambda \bar{\Lambda} \eta}$)  & $(2.26\pm0.30\pm0.20)\times10^{-5}$& $(5.72\pm1.34\pm0.65)\times10^{-6}$& $(1.00\pm0.20\pm0.14)\times10^{-5}$\\
  $\mathcal{B}({\psi(3686)\to\gamma\chi_{cJ}})$~\cite{pdg2020}  & $(9.79\pm0.20)\%$ & $(9.75\pm0.24)\%$ & $(9.52\pm0.20)\%$\\
 $\mathcal{B}({\chi_{cJ}\to \Lambda \bar{\Lambda} \eta}$) & $(2.31\pm0.30\pm0.21)\times10^{-4}$& $(5.87\pm1.38\pm0.68)\times10^{-5}$& $(1.05\pm0.21\pm0.15)\times10^{-4}$\\   \hline
\hline
\end{tabular}
\end{table*}

\section{SYSTEMATIC UNCERTAINTIES}

\label{sec:systematics}

The systematic uncertainties in the branching fraction measurements,
described below, come from several sources, as summarized in
Table~\ref{tab:systematics}.

\begin{table}[htbp]
\caption{Systematic uncertainties~(\%) in the measurements of the branching fractions of $\chi_{cJ}\to \Lambda\bar \Lambda\eta$. }
   \centering
   \begin{tabular}{lccc}
   \hline
   \hline
  Source  & $\chi_{c0}$& $\chi_{c1}$& $\chi_{c2}$\\
        \hline
  $N_{\psi(3686)}$                               & 0.7 & 0.7  & 0.7 \\
  $\Lambda(\bar\Lambda)$ reconstruction          & 1.8 & 5.2  & 5.7 \\
  $\gamma$ selection                             & 3.0 & 3.0  & 3.0 \\
  $\eta$ mass window                             & 1.0 & 1.0  & 1.0 \\
Rejection of  $J/\psi\to\Lambda\bar{\Lambda}\gamma$  & ... & 0.9  & 8.1 \\
Rejection of  $\chi_{cJ}\to\Lambda\bar{\Lambda}\pi^0$  & ... & 1.8  & 2.9 \\
Rejection of  $\Sigma(\bar{\Sigma})$             & ... & 6.3  & 4.5 \\
  $M_{\Lambda\bar{\Lambda}\eta}$ fit             & 6.8 & 4.0  & 2.1 \\
  BODY3 generator                                & 3.2 & 4.7  & 7.5 \\
  4C kinematic fit                               & 2.0 & 2.3  & 2.8 \\
  MC statistics                                  & 0.7 & 0.6  & 0.6 \\
 $\mathcal{B}({\psi(3686) \to \gamma\chi_{cJ}})$  & 2.0 & 2.5  & 2.1 \\
 $\mathcal{B}({\Lambda \to p\pi^-})$              & 1.6 & 1.6  & 1.6 \\
 $\mathcal{B}({\eta\to\gamma\gamma})$             & 0.5 & 0.5  & 0.5 \\
 \hline
   Total                                         & 9.0 &11.6 &14.6 \\
   \hline
   \hline
  \end{tabular}
  \label{tab:systematics}
\end{table}

The systematic uncertainty of the $\psi(3686)$ event number $N_{\psi(3686)}$ is 0.7\%~\cite{ref::psip-num-inc}.

The efficiencies of $\Lambda(\bar{\Lambda})$ reconstruction, including
the tracking efficiencies of the $p \pi^{-}(\bar{p}\pi^{+})$ pair,
decay length requirement, mass window requirement, vertex fit and
second vertex fit, are studied using the control samples of $J/\psi\to
pK^-\bar \Lambda+c.c.$ and $J/\psi\to\Lambda\bar{\Lambda}$. The
efficiency difference between data and MC simulation $\Delta \epsilon^{i}=\epsilon_{i}^{\rm data}/\epsilon_{i}^{\rm MC}-1$ in each momentum and $\cos\theta$ bin of data is
weighted by $w_i = \frac{N_i}{N_{\rm tot}}$,
where $N_i$ is the number of generated MC events in the $i$-th bin
and $N_{\rm tot}$ is the total number of generated MC  events.
The differences of the detection efficiencies
between data and MC simulation $\sum w_i\times\Delta \epsilon^{i}$ are assigned as the corresponding systematic
uncertainties, which are 1.6\%, 2.9\%, and 3.0\% for $\Lambda$ and 0.2\%,
2.3\%, and 2.7\% for $\bar{\Lambda}$ in $\chi_{c0}\to \Lambda\bar
\Lambda \eta$, $\chi_{c1}\to \Lambda\bar \Lambda \eta$, and
$\chi_{c2}\to \Lambda\bar \Lambda \eta$, respectively.

The systematic uncertainty due to the photon detection is determined
to be 1.0\% per photon by using the control sample
$J/\psi\to\pi^+\pi^-\pi^0$~\cite{ref::gamma-recon}.

The systematic uncertainty related to the $\eta$ mass window is studied with the control sample of
$\psi(3686)\to\eta J/\psi, J/\psi\to l^+l^-$ $(l=e,\mu)$.
The difference of the acceptance efficiencies between data and MC simulation, 1.0\%, is
taken to be the corresponding systematic uncertainty.

The systematic uncertainties arising from rejections of $J/\psi\to\Lambda\bar{\Lambda}\gamma$,
$\chi_{cJ}\to\Lambda\bar{\Lambda}\pi^0$, and $\Sigma(\bar{\Sigma})\to\Lambda(\bar{\Lambda})\gamma$
are estimated by varying individual rejection windows by one time of resolutions
in $M_{\Lambda\bar{\Lambda}\gamma}$, $M_{\gamma\gamma}$, and
$M_{\Lambda(\bar{\Lambda})\gamma}$, respectively. Totally 1000 pseudo-data-sets are sampled with replacement data for each case according to the bootstrap method~\cite{ref::bootstrap}. For each pseudo-data set, similar fit is performed on $M_{\Lambda\bar \Lambda\eta}$ as the fit to real data. We examine the pull distribution relative to the fit yield of real data, which is
\begin{equation}
	p(N_{\rm sig})=\frac{N^{\rm pseudo}_{\rm sig}-N^{\rm real}_{\rm sig}}{\sigma_{N^{\rm pseudo}_{\rm sig}}},
\end{equation}
where $N^{\rm real}_{\rm sig}$ is the fit yield of real data, and $N^{\rm pseudo}_{\rm sig}$ and ${\sigma_{N^{\rm pseudo}_{\rm sig}}}$ are the fit yield and its statistical uncertainty of pseudo-data, respectively. We fit to this pull distribution with a Gaussian function. If the mean value is larger than two times of its standard deviation, the maximum deviation will be assigned as the corresponding systematic uncertainty. The systematic uncertainties due to $J/\psi\to\Lambda\bar{\Lambda}\gamma$, $\chi_{cJ}\to\Lambda\bar{\Lambda}\pi^0$, and $\Sigma(\bar{\Sigma})\to\Lambda(\bar{\Lambda})\gamma$ rejections are assigned to be 0.9\%, 1.8\%, and 6.3\% for $\chi_{c1}\to\Lambda\bar{\Lambda}\eta$; and 8.1\%, 2.9\%, and 4.5\% for $\chi_{c2}\to\Lambda\bar{\Lambda}\eta$, respectively; while those for $\chi_{c0}\to\Lambda\bar{\Lambda}\eta$ are negligible with deviation less than 2$\sigma$.

The systematic uncertainties in the $M_{\Lambda\bar{\Lambda}\eta}$ fit are considered in two aspects. The systematic uncertainties associated with the
signal shape are estimated by using alternative signal shapes based on MC simulation. The
changes of the fitted signal yields, 2.2\%, 2.3\%, and 1.9\% for
$\chi_{c0}\to \Lambda\bar{\Lambda}\eta$, $\chi_{c1}\to
\Lambda\bar{\Lambda}\eta$, and $\chi_{c2}\to
\Lambda\bar{\Lambda}\eta$, respectively, are taken as the
corresponding systematic uncertainties.  The systematic uncertainties from the
background shape are estimated by using a second order polynomial. The
changes of the fitted signal yields, 6.4\%, 3.3\%, and 1.0\% for
$\chi_{c0}\to \Lambda\bar{\Lambda}\eta$, $\chi_{c1}\to
\Lambda\bar{\Lambda}\eta$, and $\chi_{c2}\to
\Lambda\bar{\Lambda}\eta$, respectively, are taken as the
corresponding systematic uncertainties.  Adding them in quadrature,
we obtain the systematic uncertainties due to the $M_{\Lambda\bar{\Lambda}\eta}$ fit to be 6.8\%, 4.0\%, and 2.1\%
for $\chi_{c0}\to \Lambda\bar{\Lambda}\eta$, $\chi_{c1}\to
\Lambda\bar{\Lambda}\eta$, and $\chi_{c2}\to
\Lambda\bar{\Lambda}\eta$, respectively.

The systematic uncertainty due to the BODY3 generator is estimated by varying the weight in each bin by $\pm 1\sigma$.
The weights in various bins are obtained with data after subtracting the normalized background from inclusive $\psi(3686)$ MC sample.
The change of the weighted signal efficiency caused by each bin is obtained to be $\Delta\epsilon_i$.
Summing $\Delta\epsilon_i$ over all bins with $\sqrt{\sum_{i=1}^{n_{\rm bin}}\Delta\epsilon_i^2}$,
the associated systematic uncertainties are obtained to be 3.4\%, 4.7\%, and  7.5\%  for $\chi_{c0}\to \Lambda\bar{\Lambda}\eta$,
$\chi_{c1}\to \Lambda\bar{\Lambda}\eta$, and
$\chi_{c2}\to \Lambda\bar{\Lambda}\eta$, respectively.

The systematic uncertainties of the 4C kinematic fit are assigned as
the differences between the detection efficiencies before and after the helix
parameter corrections \cite{ref::helix-correction}, which are
2.0\% for $\chi_{c0}\to\Lambda\bar{\Lambda}\eta$, 2.3\% for
$\chi_{c1}\to\Lambda\bar{\Lambda}\eta$, and 2.8\% for
$\chi_{c2}\to\Lambda\bar{\Lambda}\eta$.

The uncertainties due to the limited MC statistics are calculated
from $\sqrt{\frac{ 1-\epsilon }{N\cdot\epsilon}}$,
where $\epsilon$ is the detection efficiency and $N$ is the
number of signal MC events. They are 0.7\%, 0.6\%, and 0.6\% for $\chi_{c0}\to
\Lambda\bar \Lambda \eta$, $\chi_{c1}\to \Lambda\bar \Lambda \eta$, and
$\chi_{c2}\to \Lambda\bar \Lambda \eta$, respectively.

The uncertainties from the world averages of the branching fractions of
$\psi(3686)\to\gamma\chi_{c0}$, $\psi(3686)\to\gamma\chi_{c1}$, $\psi(3686)\to\gamma\chi_{c2}$,
$\Lambda \to p\pi^-$, and $\eta\to\gamma\gamma$~\cite{pdg2020} are 2.0\%, 2.5\%, 2.1\%, 0.8\%, and 0.5\%, respectively.

For each signal decay, the total systematic uncertainty is calculated
by adding systematic uncertainties quadratically under the assumption
that all sources are independent.

\section{Summary}

By analyzing $(448.1\pm2.9)\times10^6$ $\psi(3686)$ events collected
with the BESIII detector, we observe the decays of $\chi_{c0,1,2}\to
\Lambda\bar \Lambda \eta$ for the first time.  The product branching
fractions of $\psi(3686)\to\gamma\chi_{cJ}$ and $\chi_{cJ}\to
\Lambda\bar \Lambda \eta$ are determined. Dividing by the world
averages of the branching fractions of $\psi(3686)\to\gamma\chi_{cJ}$,
we determine the branching fractions of $\chi_{cJ}\to \Lambda\bar
\Lambda \eta$ as summarized in Table \ref{tab:Branching}.
The current available statistics is not sufficient to draw any conclusion that there is an enhancement near the
$\Lambda\bar{\Lambda}$ production. We looked at the $M_{\Lambda \eta}$ or $M_{\bar{\Lambda}\eta}$ spectra and did not find any excited $\Lambda$ state.


\section{ACKNOWLEDGMENTS}
The BESIII collaboration thanks the staff of BEPCII and the IHEP computing center for their strong support. This work is supported in part by National Key R\&D Program of China under Grants No. 2020YFA0406300 and No. 2020YFA0406400; National Natural Science Foundation of China (NSFC) under Grants No. 12035009, No. 11875170, No. 11475090, No. 11635010, No. 11735014, No. 11835012, No. 11935015, No. 11935016, No. 11935018, No. 11961141012, No. 12022510, No. 12025502, No. 12035013, No. 12192260, No. 12192261, No. 12192262, No. 12192263, No. 12192264 and No. 12192265; the Chinese Academy of Sciences (CAS) Large-Scale Scientific Facility Program; Joint Large-Scale Scientific Facility Funds of the NSFC and CAS under Grant No. U1832207; CAS Key Research Program of Frontier Sciences under Grant No. QYZDJ-SSW-SLH040; 100 Talents Program of CAS; The Institute of Nuclear and Particle Physics (INPAC) and Shanghai Key Laboratory for Particle Physics and Cosmology; ERC under Grant No. 758462; European Union's Horizon 2020 research and innovation programme under Marie Sklodowska-Curie grant agreement under Grant No. 894790; German Research Foundation DFG under Grants No. 443159800, Collaborative Research Center CRC 1044, GRK 2149; Istituto Nazionale di Fisica Nucleare, Italy; Ministry of Development of Turkey under Grant No. DPT2006K-120470; National Science and Technology fund; National Science Research and Innovation Fund (NSRF) via the Program Management Unit for Human Resources \& Institutional Development, Research and Innovation under Grant No. B16F640076; STFC (United Kingdom); Suranaree University of Technology (SUT), Thailand Science Research and Innovation (TSRI), and National Science Research and Innovation Fund (NSRF) under Grant No. 160355; The Royal Society, UK under Grants No. DH140054 and No. DH160214; The Swedish Research Council; U. S. Department of Energy under Grant No. DE-FG02-05ER41374.

\end{document}